\documentclass{acm_proc_article-sp}

\usepackage{url}
\usepackage{multirow}
\usepackage{rotating}
\usepackage{bbm}
\usepackage{amsmath}
\usepackage{graphicx}

\begin{document}

\title{Social Network Dynamics in a Massive Online Game: \\ Network Turnover, Non-densification, and Team Engagement in Halo Reach}

\numberofauthors{2} 
\author{
\alignauthor
Sears Merritt\\
       \affaddr{University of Colorado Boulder}\\
       \email{sears.merritt@colorado.edu}
\alignauthor
Aaron Clauset\\
        \affaddr{University of Colorado Boulder}\\
        \affaddr{Santa Fe Institute}\\
        \email{aaron.clauset@colorado.edu}
}

\maketitle

\begin{abstract}
Online multiplayer games are a popular form of social interaction, used by hundreds of millions of individuals. However, little is known about the social networks within these online games, or how they evolve over time. 
Understanding human social dynamics within massive online games can shed new light on social interactions in general and inform the development of more engaging systems. Here, we study a novel, large friendship network, inferred from nearly 18 billion social interactions over 44 weeks between 17 million individuals in the popular online game \textit{Halo:\!\! Reach}. This network is one of the largest, most detailed temporal interaction networks studied to date, and provides a novel perspective on the dynamics of online friendship networks, as opposed to mere interaction graphs. 
Initially, this network exhibits strong structural turnover and decays rapidly from a peak size. In the following period, however, both network size and turnover stabilize, producing a dynamic structural equilibrium. In contrast to other studies, we find that the Halo friendship network is non-densifying: both the mean degree and the average pairwise distance are stable, suggesting that densification cannot occur when maintaining friendships is costly. Finally, players with greater long-term engagement exhibit stronger local clustering, suggesting a group-level social engagement process. These results demonstrate the utility of online games for studying social networks, shed new light on empirical temporal graph patterns, and clarify the claims of universality of network densification.
\end{abstract}




\section{Introduction}
Although social networks are inherently dynamic, relatively few studies have analyzed the dynamical patterns observed in large, real-world, online social networks. Insights into the basic organizing principles and patterns of these networks should inform the development of probabilistic models of their evolution, the development of novel methods for detecting anomalous dynamics, as well as the design of novel, or augmentation of existing, online social systems so as to support better user engagement. 
Here, we present a brief empirical analysis of a novel online social network, and use its dynamics as a lens by which to both identify interesting empirical patterns and test existing claims about social network dynamics.

This network is drawn from a massive online game \textit{Halo:\!\! Reach}, which includes the activities of 17 million unique individuals across roughly 2,700,000 person-years of continuous online interaction time. Multiplayer games like these are a highly popular form of online interaction, and yet, likely due to the lack of data availability, have rarely been studied in the context of online social networks. As a result, we know more about networks like Facebook and Twitter, which consume relatively little time per user per week, than we do about more immersive online social systems like multiplayer games, which consume on average more than 20 hours of time per user per week~\cite{esa2011facts}. 

Furthermore, online games represent a large economic sector, including not just the sale of the software itself, but an entire ecosystem of entertainment products, worth billions of US dollars worldwide.
For example, on the first day of sale \textit{Halo:\!\! Reach} grossed nearly \$200 million in revenue from roughly 4 million copies of the game. Increasingly, one point of attraction to this and other online games is their multiplayer component: while it is possible to play individually, the online system is designed to encourage and reward social play. Despite their enormous popularity, relatively little is known about the social networks within these online games, how they evolve over time, and whether the patterns they exhibit agree or disagree with what we know about online social interactions from other online systems.

This raises two interesting questions. First, what is the shape of a social network in a massive, multiplayer online game? And second, how does this network change over time?  Answers to these questions will shed new light on the relative importance of a game's social dimension and on the relationship between a game's social structure and its long term success, and will clarify the generality of our current beliefs about online social networks, which are largely derived from non-game systems like Facebook, Twitter, Flickr, etc.

One immediate difference between the social network underlying online games and those more classically studied is the competitive nature of games. The large scale of and rich data produced by online games provides a novel perspective on certain other social interactions, e.g., team competition. Past work has shed light on competitive dynamics~\cite{merritt2013environmental}, social organization~\cite{szell2010multirelational}, economic trading networks~\cite{keegan2010dark}, and deviant behavior~\cite{blackburn2012branded}. Here, we study the underlying friendship network from \textit{Halo:\!\! Reach}, which we infer~\cite{merritt2013detecting} from 18 billion interactions between 17 million individuals over the course of 44 weeks. Edges in this network represent inferred online or offline friendships, in contrast to mere online interactions alone, which the Halo system generates randomly via the matchmaking system that places players into competitions.

We first analyze the static (cumulative) friendship network and find that it contains a heavy-tailed degree distribution, which appears more log-normal than power law. This network is composed of many small disconnected components and a single giant component containing roughly 30\% of all players. A plurality of vertices exhibit low clustering coefficients ($0< c_{i} \leq 0.1$), indicating very sparse local structure, while the next-most-common pattern is to exhibit a very high clustering coefficient ($0.9<c_{i}<1.0$), indicating very dense local structure. 

To study the friendship network's evolution we create a time series of 44 network snapshots, one for each week of the year. Initially, this network exhibit strong structural turnover and decays rapidly from a peak size to a more stable, but dynamic core. Despite its apparently stable size, we find steady network turnover over most of the time period, indicating a dynamic equilibrium as roughly equal numbers of vertices join and leave the network each week. Furthermore, in contrast to other online social networks~\cite{Kumar:2006}, the Halo friendship network does not densify over time: the mean degree and the average pairwise distance within the giant component are stable.

Finally, we observe that the network's clustering grows over the first 25 weeks and then declines, suggesting that players form increasingly tight knit groups during the first half of the network's life only to disband them later. In addition, we find a positive relationship between the consecutive weeks played and the player's mean clustering coefficient. This indicates that players who belong to tightly knit groups tend to also play longer and more consistently.

The remainder of this paper is organized as follows: first we discuss related work. Next, for completeness, we introduce the data and the results of an anonymous online survey. Following this, we briefly describe the method used to infer the social network. These topics are described in detail in~\cite{merritt2013detecting}. Following this, we analyze the inferred social network statically and then dynamically. We conclude with a discussion and final thoughts on future work.

\vspace{1cm}

\section{Related work}
Work related to the analysis of networks began with the study of static networks. Researchers identified important structural patterns such as heavy tailed degree distributions~\cite{barabasi1999emergence}, small-world phenomenon~\cite{milgram1967small,watts1998collective}, and communities, as well as probabilistic models and algorithms that produce and detect them~\cite{newman2003structure,clauset2004finding}.

As new online social systems continue to emerge on the web, the static analysis of social networks continues to be an area of great interest. Researchers have a steady stream of new empirical network data with which they can test new and existing theories about social dynamics, whose sources include Twitter~\cite{kwak2010twitter}, Facebook~\cite{ugander2011anatomy}, Orkut and Flicker~\cite{mislove2007measurement}.

More recently, researchers have begun to study how time influences~\cite{clauset2007persistence} and changes network structure~\cite{ahn2007analysis}. New dynamical patterns, such as densification~\cite{leskovec2005graphs,Kumar:2006} and shrinking $k$-cores~\cite{garcia2013social}, as well as probabilistic models have been identified that shed light on how changes in the underlying processes that produce these networks affect its structure.

Since the underlying processes that produce complex networks are typically not random, a rich body of work related to mathematically modeling various networks whose structure cannot be generated by an Erd\H{o}s-R\'enyi random graph model has emerged~\cite{erdds1959random}. One such model is the configuration model, which produces a graph with a predefined degree distribution by randomly assigning edges between vertices according to their degree sequence~\cite{newman2001random}. Another is the preferential attachment model, which is a generative model that produces heavy tailed degree distributions by assigning edges to vertices according to the notion of ``the rich get richer". That is, edges are assigned to vertices according to how many they already have. Other approaches that produce specific properties, such as short diameters and communities include the Watts and Strogatz model and stochastic block model respectively~\cite{watts1998collective,girvan2002community}. 

Lastly, we would be remiss if we did not also mention work related to the study of online games. Most uses of online game data have focused on understanding certain aspects of human social behavior in online environments. Examples include individual and team performance~\cite{shim2010player,shim2010team,shim:etal:2011,shim2011exploratory}, expert behavior~\cite{huffaker2009social,huang2013mastering}, homophily~\cite{huang2009virtually}, group formation~\cite{huang2009formation}, economic activity~\cite{castronova2009real,bakshy2010social}, and deviant behavior~\cite{ahmad2009mining}. Most of this work has focused on massively multiplayer online role playing games (MMORPGs), e.g., World of Warcraft, although a few have examined social behavior in first person shooter (FPS) games like \textit{Reach}~\cite{shim:etal:2011}. Relatively little of this work has focused on the structure and dynamics of social networks. 

A particularly unique aspect of our work is our data set. Few studies have analyzed the temporal dynamics of a social network derived from a popular form of social media, online games. Moreover, the majority of studies that examine the temporal dynamics of networks only analyze networks that primarily grow over time. That is, once a vertex is added to these networks it is never removed. In our network, vertices enter and leave the network consistently over time.

\vspace{1cm}

\section{Data and survey}
\subsection{Game details}
Our data are derived from detailed records of game play from \textit{Halo:\!\! Reach}, a popular online first person shooter game. Individual game files were made available through the Halo Reach Stats API.\footnote{The API was active from September 2010 through November 2012. API documentation was taken offline in September 2012.} Through this interface, we collected the first 700 million game instances (roughly 305 days of activity by 17 million individuals). Among other information, each game file includes a Unix timestamp, game type label, and a list of gamertags. This large database provides us with complete data on the timing and character of interactions between individuals but provides no information about which interactions are produced by friendships versus non-friendships.

\begin{figure}[t!]
\begin{center}
\includegraphics[scale=0.4]{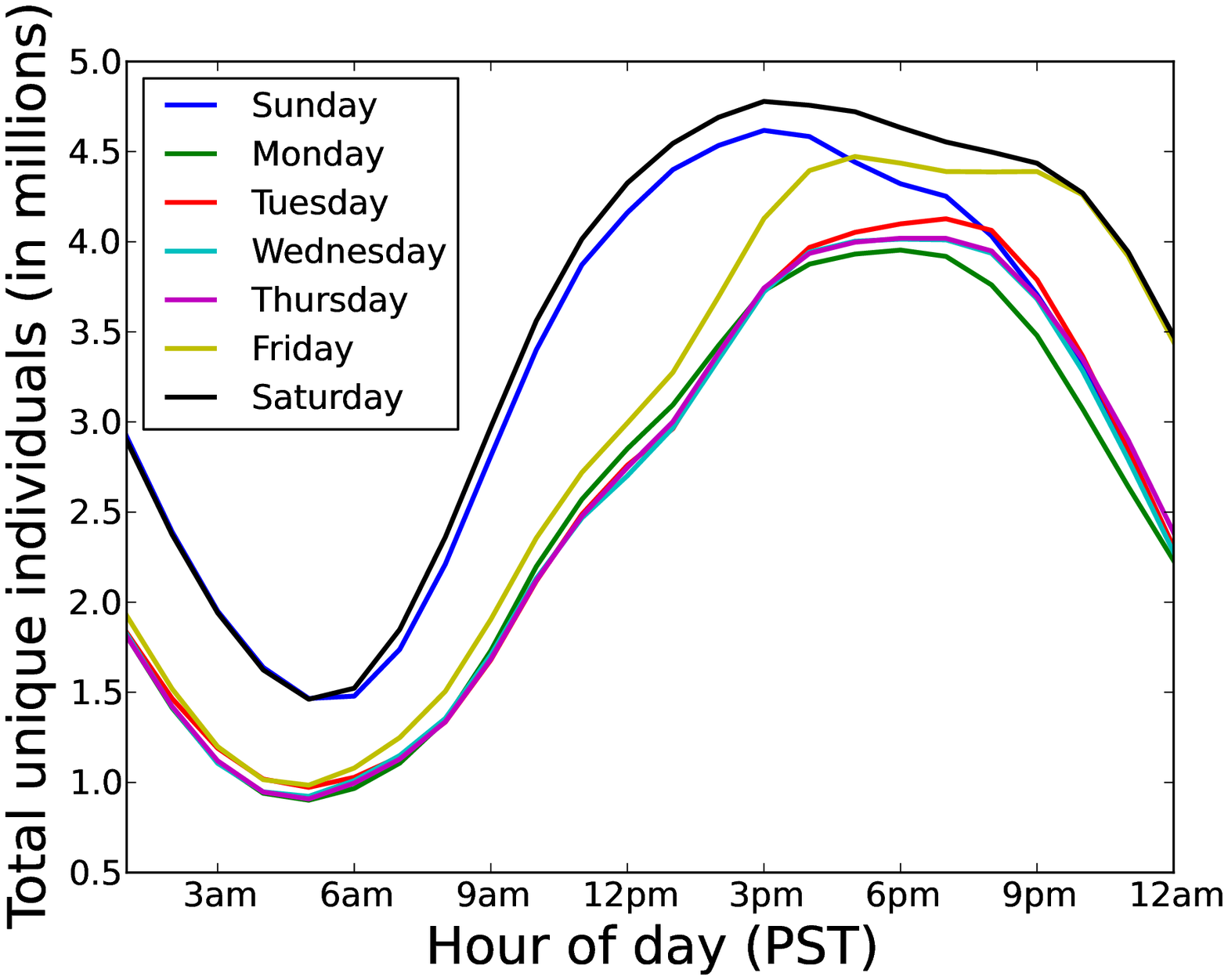} \\
\includegraphics[scale=0.4]{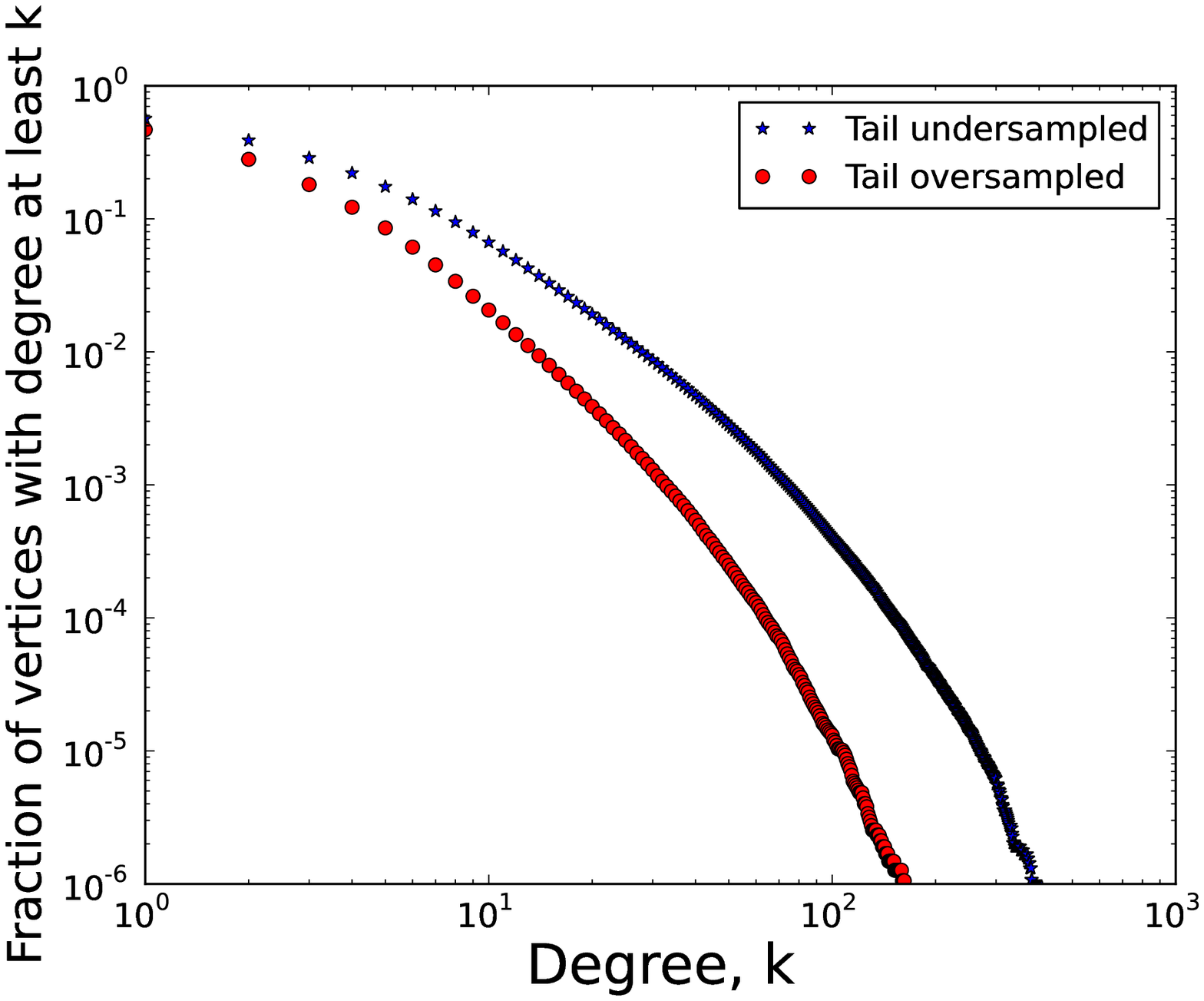}
\caption{(Top) Number of unique individuals ever seen at a given time of day (in Pacific Standard Time), across the 305 days spanned by the data, illustrating significant daily and weekly periodicities. (Bottom) Complementary cumulative distribution functions of vertex degrees}
\label{fig:players:per:hour}
\end{center}
\end{figure}

\vspace{1cm}
\subsection{Survey}
We combine these in-game behavioral data with the results of an anonymous online survey of \textit{Reach} players~\cite{mason2012friends}. In the survey, participants supplied their gamertag from which we generated a list of all other gamertags that had ever appeared in a game with the participant. From this list, the participant identified which individuals were friends.~\footnote{In the survey, a friend is defined as a person known by the respondent at least casually, either offline or online.}  We interpret these subjective friendship labels as ground truth. From these data, we constructed a social network with links pointing from participants to their labeled friends. In our supervised learning analysis, both a labeled friendship and the absence of a label are treated as values to be predicted (i.e., we assume survey respondents explicitly chose not to label their co-player as a friend). Of the 965 participants who had completed the friendship portion of the survey by April 2012, 847 individuals appear in our data (the first 305 days of play); this yielded 14,045 latent friendship ties and 7,159,989 non-friendship ties.

\subsection{Interaction network}
We represent the set of pairwise interactions as a temporal network, in which edges have endpoints and exist at a specific moment in time. Vertices in the network correspond to gamertags, and two vertices are connected if they appear in a game instance together at time $t$ (time of day, in 10 minute intervals). Each vertex thus has a sequence or time series of interactions with other vertices. The resulting network, derived from our complete game sample, contains 17,286,270 vertices, 18,305,874,864 temporal edges, and spans $305$ days. The subgraph of interactions by our survey participants contained a total of 2,531,479 vertices and 665,401,283 temporal edges over the same period of time.

\begin{figure*}[t!]
\begin{center}
\begin{tabular}{ccc}
\includegraphics[scale=0.28]{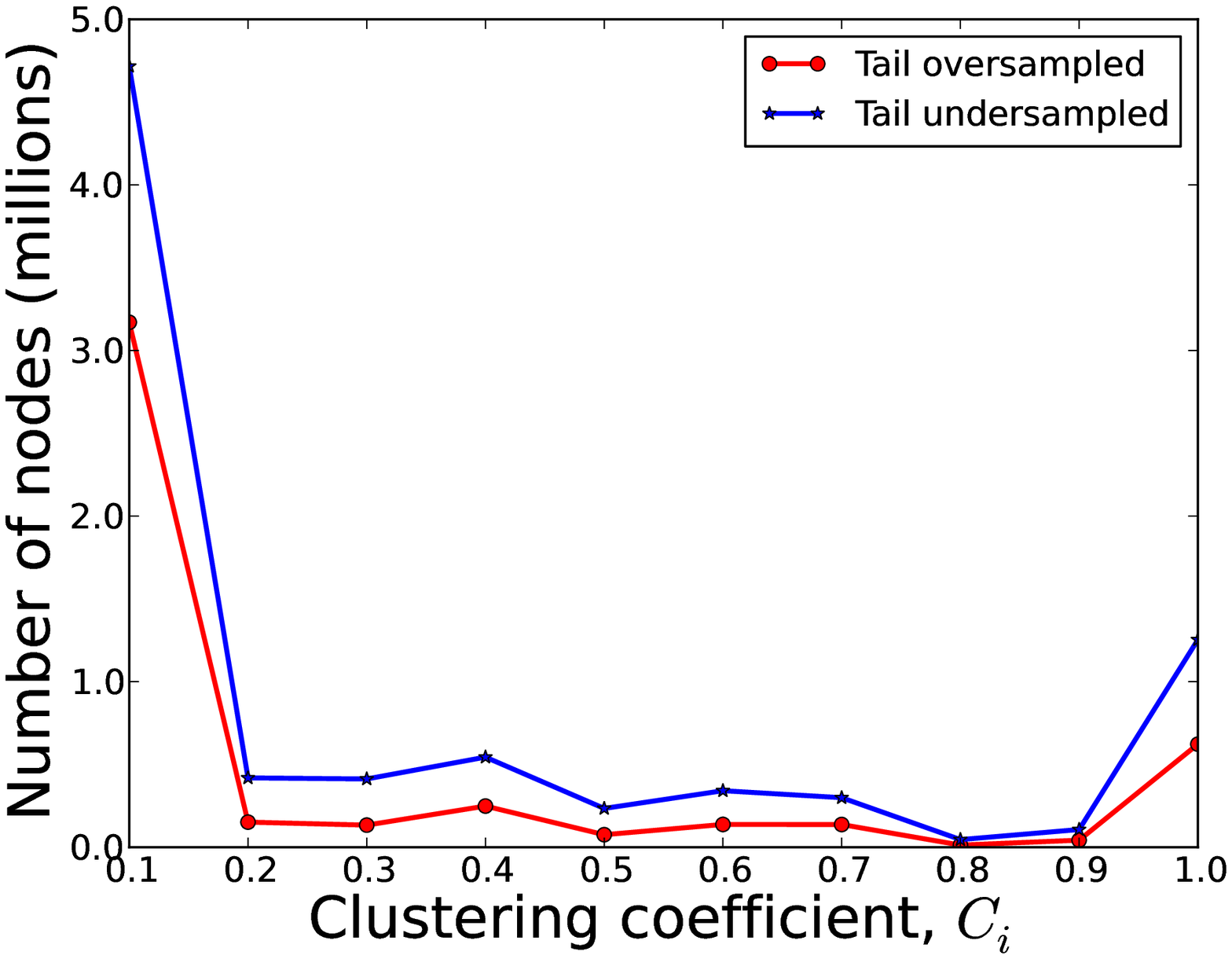} &
\includegraphics[scale=0.28]{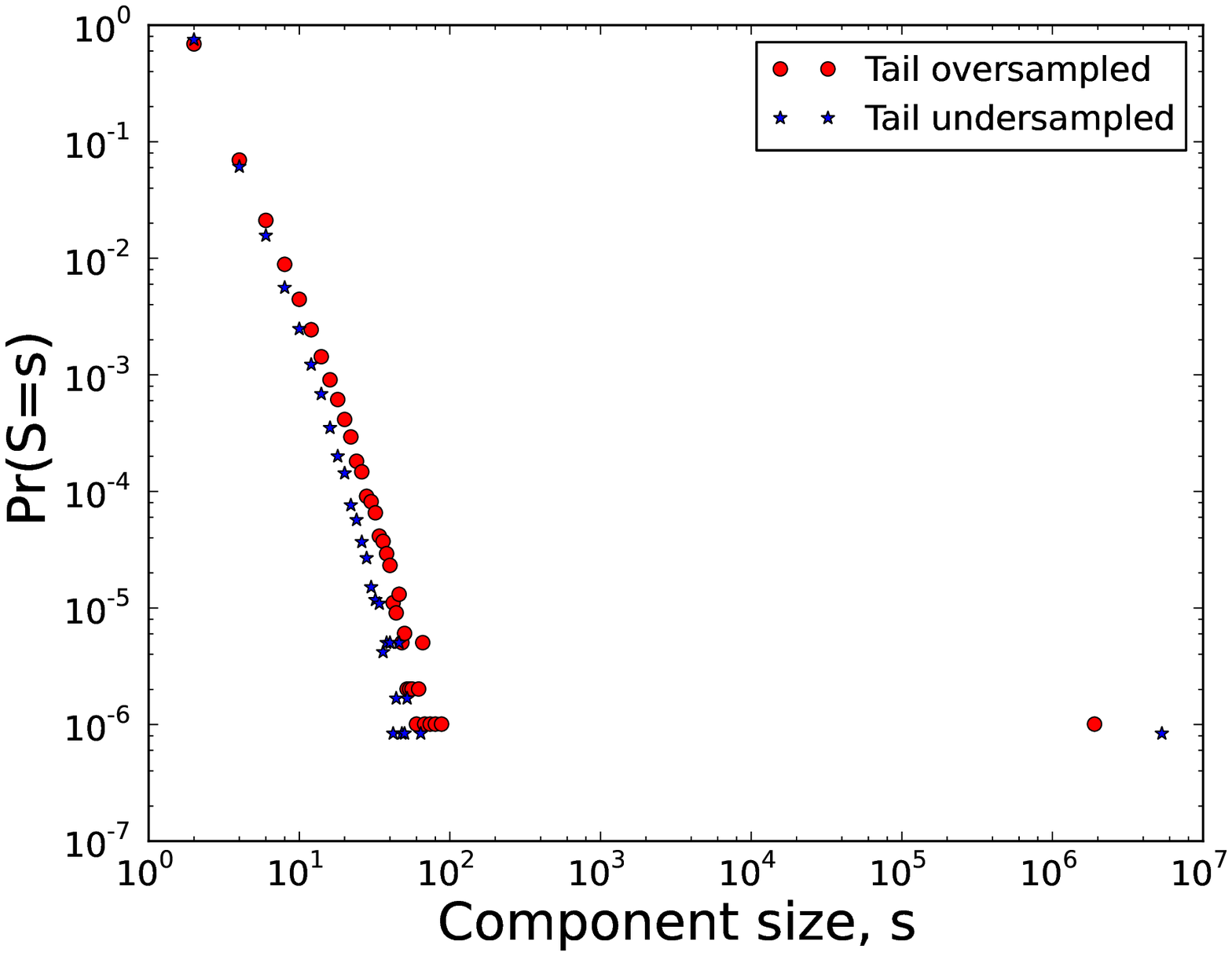} &
\includegraphics[scale=0.28]{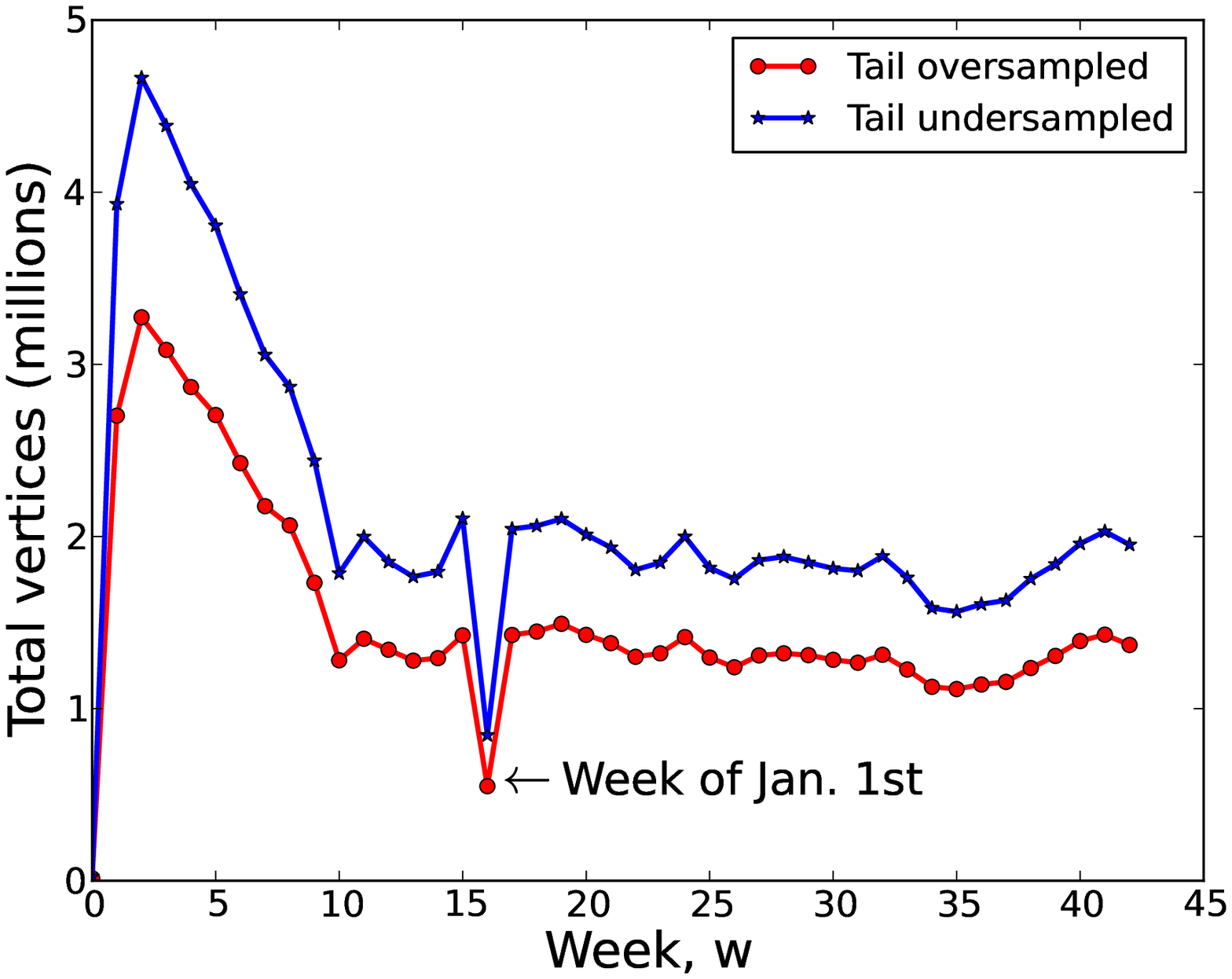}
\end{tabular}
\caption{(Left) Histogram of vertex level clustering coefficients, $C_i$ (Center) Distribution of component sizes. (Right) Total number of vertices in the network as a function of week, $w$. All figures plot distributions for networks using under and over sampled tail thresholds ($AC_{x,y}=197$ and $AC_{x,y}=1900$ respectively)}
\label{fig:dd:cc:cs}
\end{center}
\end{figure*}

\section{Inferring the social network}
\label{sec:prediction}
Pairs of individuals in \textit{Reach} that are friends are known to play many more consecutive games (12, on average, or about 2 hours of time) than non-friends (1.25, on average)~\cite{mason2012friends}. Thus, continuous interaction over a significant span of time is likely an indication of a latent tie, while more intermittent interactions likely indicate a non-friend tie, given the large population of non-friends available to play at any time. The expected diurnal and weekly cycles observed in the data will modulate these behaviors, and a reasonable approach for their quantification is via interaction periodicity (see Fig.~\ref{fig:players:per:hour}, top). Let
\begin{align}
n_{x,y}(t) = \mathbbm{1}\{\textrm{$x$ and $y$ play together at time $t$}\}
\end{align}
represent the time series of binary interactions between individuals $x$ and $y$, where $1$ indicates an interaction at time $t$ and $0$ indicates no interaction. If $x$ and $y$ are friends, we expect $n_{x,y}(t)$ to exhibit stronger periodicity than for non-friends. This expectation may be quantified as the autocorrelation of the time series $n_{x,y}(t)$ over all time lags $\tau$:
\begin{align}
AC_{x,y} & = \sum_\tau \sum_t n_{x,y}(t)n_{x,y}(t-\tau).
\end{align}
If $n_{x,y}(t)$ is generated by a non-friend pair, $AC_{x,y}$ should be small because these individuals do not interact regularly. On the other hand, if $n_{x,y}(t)$ is generated by a friend pair, we expect $AC_{x,y}$ to be large. Training a logistic regression classifier on this feature alone correctly identifies 95\% of ties, even for casual users (i.e. those that play few games). For a detailed presentation of this analysis, see~\cite{merritt2013detecting}.

The survey respondents are a biased sample of \textit{Reach} players~\cite{mason2012friends}, being substantially more skilled than the typical player and investing roughly an order of magnitude more time playing than an average player. It is thus possible that the survey sampling bias has produced an oversampling or an undersampling of the tail of the degree distribution.

In an attempt to control these opposing biases, we choose two thresholds, one to show what the network looks like if the survey respondents have less friends (undersampled tail) than the population, where $AC_{x,y}=197$, and one to show network structure if the respondents have more (oversampled tail), $AC_{x,y}=1900$. Details of the methods used to select these thresholds can be found in~\cite{merritt2013detecting}.

\section{Network structure}
\label{sec:structure}
The two thresholds computed using the survey data in the previous section represent reasonable bounds on what we expect to observe in the data at large. In this section, we use both thresholds to analyze the structure of the inferred social network and show that network structure remains invariant to threshold choice. In the undersampled tail scenario ($AC_{x,y}=197$), the inferred network consists of 8,373,201 nodes and 31,051,991 edges, while the network inferred using the oversampled tail threshold ($AC_{x,y}=1900$), contains 4,732,405 nodes and 11,435,351 edges.

Figure~\ref{fig:players:per:hour}(bottom) plots the complementary cumulative distribution of vertex degree sizes and indicates that the network is primarily comprised of vertices with only a few edges and only 10\% containing more than roughly 10 or 20, depending on threshold choice.

To quantitatively measure the degree to which groups make up the network, we compute the vertex level clustering coefficient, defined as,
\begin{align}
C_i = \frac{\textrm{number of connected neighbors}}{\textrm{number of possible connected neighbors}}.
\end{align}
$C_i$ provides a principled measure of how close vertex $i$ and its neighbors are to forming a clique~\cite{newman2010networks}. A clustering coefficient equal to one indicates the vertex and its neighbors form a clique, while a coefficient equal to zero indicates none of the vertex's neighbors are connected.

In addition to vertices containing only a few edges, the majority possess low clustering coefficients, indicating that most players usually choose to play games with only one other person at a time (see Fig.~\ref{fig:dd:cc:cs}, left). There is also a small, but non-trivial group of players who have high clustering coefficients, indicating that they choose to play with two or more friends, who are also friends themselves.

Figure~\ref{fig:dd:cc:cs}(center) plots the distribution of component sizes and indicates that the network contains a single large connected component composed of between two and four million players. The majority of the remaining nodes are spread amongst many components containing between roughly ten and twenty nodes. In the undersampled tail case ($AC_{x,y}=197$), the network contains 1,194,032 components. While in the oversampled tail case ($AC_{x,y}=1900$), the network contains 991,932 components. 

From this analysis, we can conclude that the network, as a whole, contains between 23\%-47\% of the total 17 million player population and that these players tend to interact with their friends in pairs or small densely connected groups.

\begin{figure*}[t!]
\begin{center}
\begin{tabular}{cccc}
\includegraphics[scale=0.205]{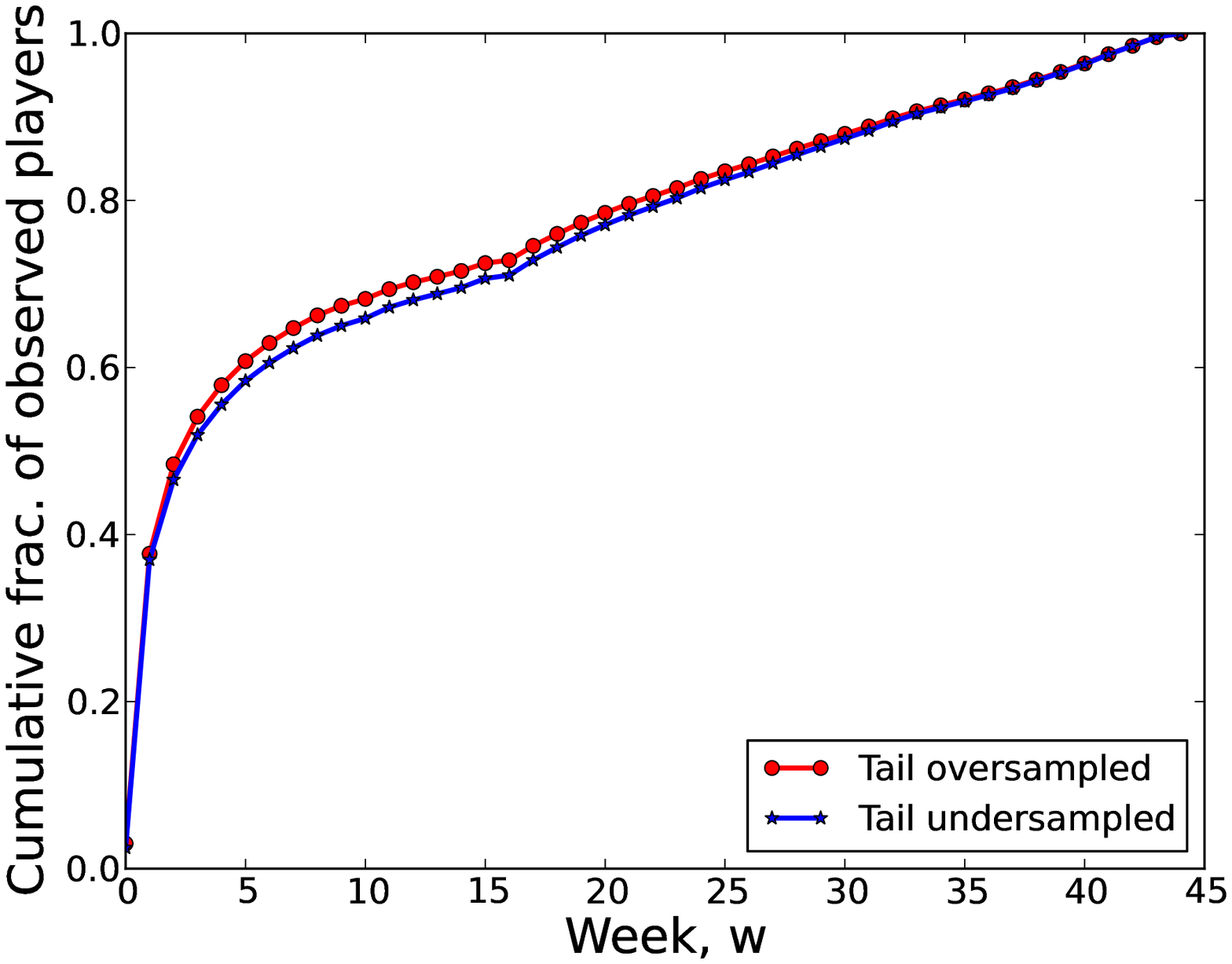} &
\includegraphics[scale=0.205]{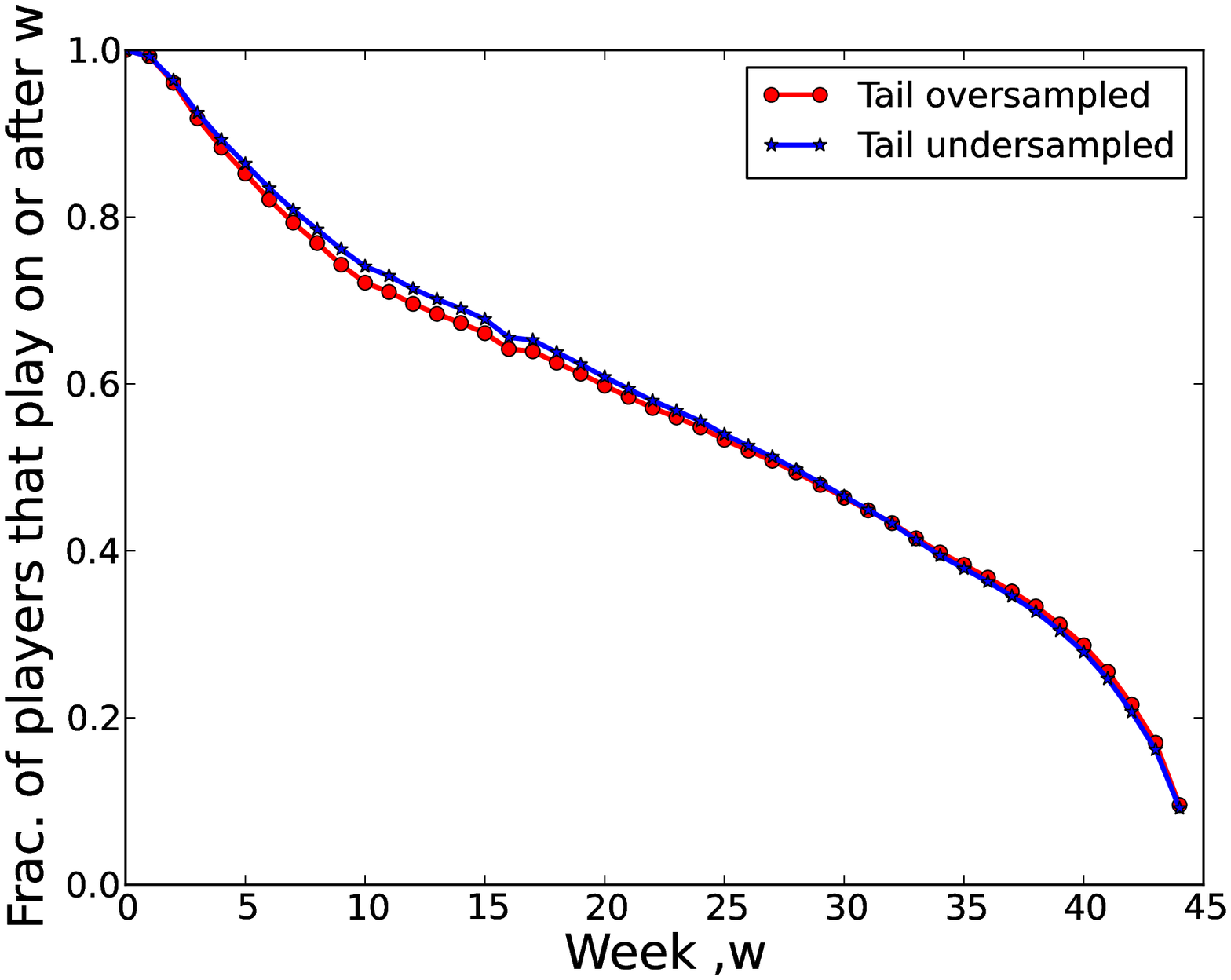} &
\includegraphics[scale=0.205]{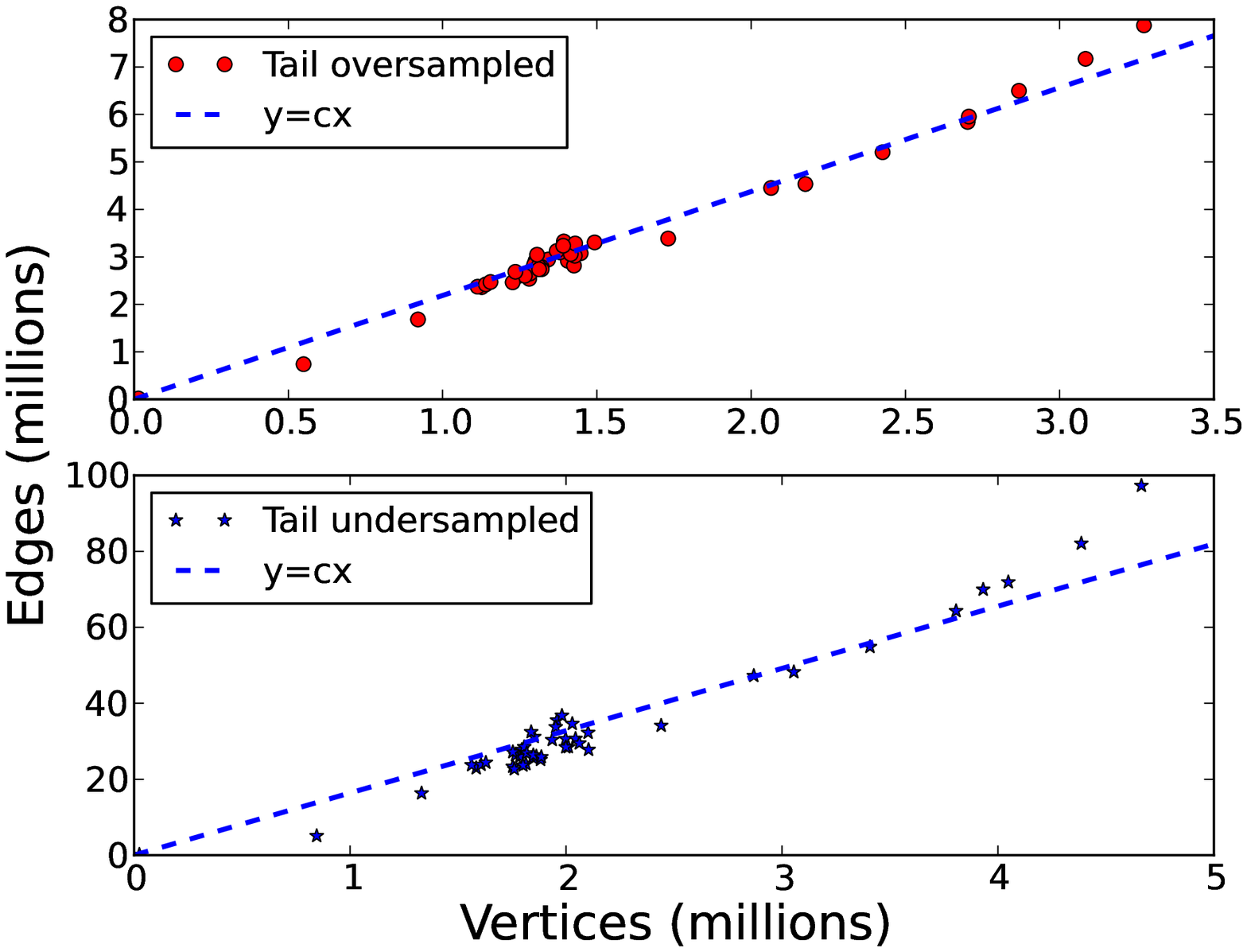} &
\includegraphics[scale=0.205]{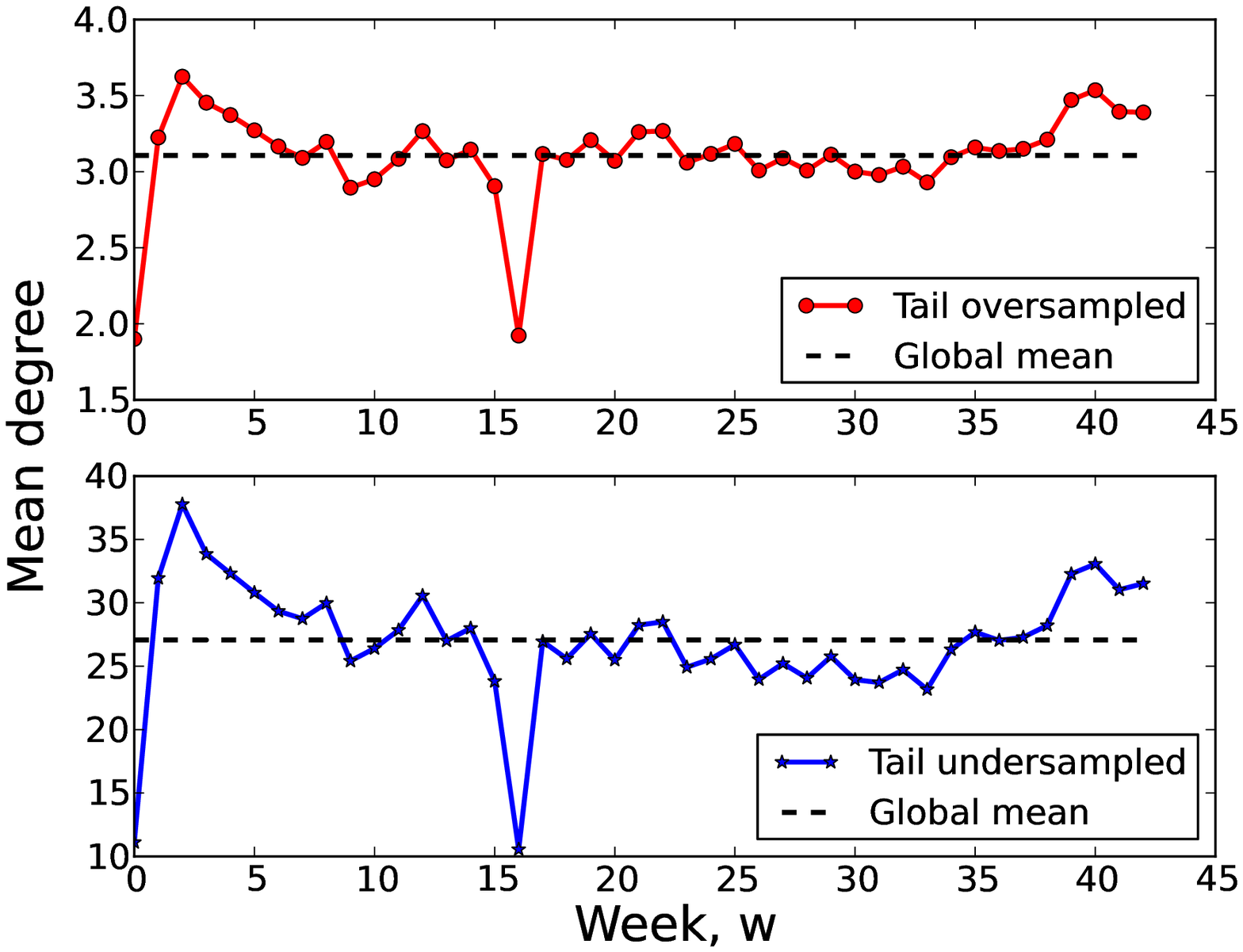}
\end{tabular}
\caption{(Left) Cumulative fraction of observed players per week, $w$. (Left-Center) Fraction of players in week $w$ that play on or after week $w$. (Right-Center) Number of edges as a function vertices and linear fit (oversampled tail $r^2=0.99$, $c=2.188$, $p \ll 0.001$, undersampled tail, $r^2=0.98$, $c=16.389$, $p \ll 0.001$), indicating that the network is \textit{not} densifying. (Right) Mean degree as a function of week, $w$}
\label{fig:vertices:per:week}
\end{center}
\end{figure*}

\section{Network dynamics}
The static analysis in the previous section sheds light on the overall structure of the social network and provides clues about how friends interact in the game, but it provides no insight into how the network changes over the course of its 44 week time span. 

To study the friendship network's evolution, we create snapshots of the network by extracting the edges of inferred friends from the interaction network at weekly time intervals, where each interval is aligned with a week of the year. The set of snapshots we study begins on the $37^{th}$ week of September, 2010, and extends through the week $28^{th}$ week of 2011, totaling 44 weeks of time.

\vspace{1cm}

\subsection{Players}
In this section, we study how the network's population evolves over time. Figure~\ref{fig:dd:cc:cs}(right) plots the inferred number of players in the social network, under both thresholds, over each week in the data. On launch week, we observe a small population that quickly grows to a peak of roughly 3.25 million and then decreases linearly to roughly 1.3 million. This pattern suggests that gamers initially play the game alone, then transition to a social mode of play{\footnote{{\em Reach} sold nearly 4 million copies of the game on the first day of sale. (see \url{http://en.wikipedia.org/wiki/Halo:_Reach})}}, and later reduce their volume of overall play. The large decrease in population can likely be attributed to players becoming bored with the game. That is, once the initial excitement and novelty of the game wears off, players spend less time playing the game.

Following the population decrease in weeks 3-10, the number of vertices remains relatively constant until week 16, the week of January $1^{st}$, where a rapid, but temporary drop, in the total population occurs. One possible cause of this drop in population could be due to an Xbox Live service outage, although we find no record of such an event on the web.

Over the course of the remaining weeks, the population slowly and mildly declines, reaching a minimum during the month of June, and then returning to the levels seen in week 10. This weak decline and growth pattern is likely due to players having to spend increasing amounts of time away from the game studying for end of semester/year exams, since the majority of players are between the ages of 16 and 24, as noted in \cite{mason2012friends}. This also explains the growth in population following the month of June, when nearly all schools have ended for the year.

\subsection{Network Turnover}
Next, we study the rates at which players enter and exit the social network over time. Figure~\ref{fig:vertices:per:week}(left) plots the cumulative fraction of observed players as a function of week, $w$. Based on the dynamics observed in Fig.~\ref{fig:dd:cc:cs}(right), it is not surprising to see that within the first 10 weeks of play, roughly 60\% of the network's population has appeared in the network at least once. One might expect the population to grow exponentially, as observed in weeks 0-10, and then become constant, indicating that a large and constant number of players enter the network early and then consistently re-appear week after week. However, the dynamics that play out following week 10 indicate that the remaining 40\% of the population continues to enter the network at a nearly constant rate. This linear growth may indeed be due to new players entering the social network, however, it may also be due, in part, to existing players changing their gamer tags, a relatively simple and inexpensive transaction (a player may change their gamertag once for free). 

The relatively constant size of the network after week 10 and the constant inflow of new players suggests that an equally sized portion of the population also exits the network over time, producing a ``dynamic equilibrium". Indeed, Figure~\ref{fig:vertices:per:week}(left-center) plots the fraction of players that reappear at least once on or after week, $w$. While the dynamics in Figure~\ref{fig:dd:cc:cs}(right) indicate players may enter the network and play a large quantity of games with friends and then exit, we find this not to be the case. The steady, nearly linear decline in the number of players that re-appear in future network snap shots after week 10 indicates that players are not quick to leave the game permanently; a pattern that is likely the result of players establishing consistent schedules with their friends.

\begin{figure*}[t]
\begin{center}
\begin{tabular}{cccc}
\includegraphics[scale=0.205]{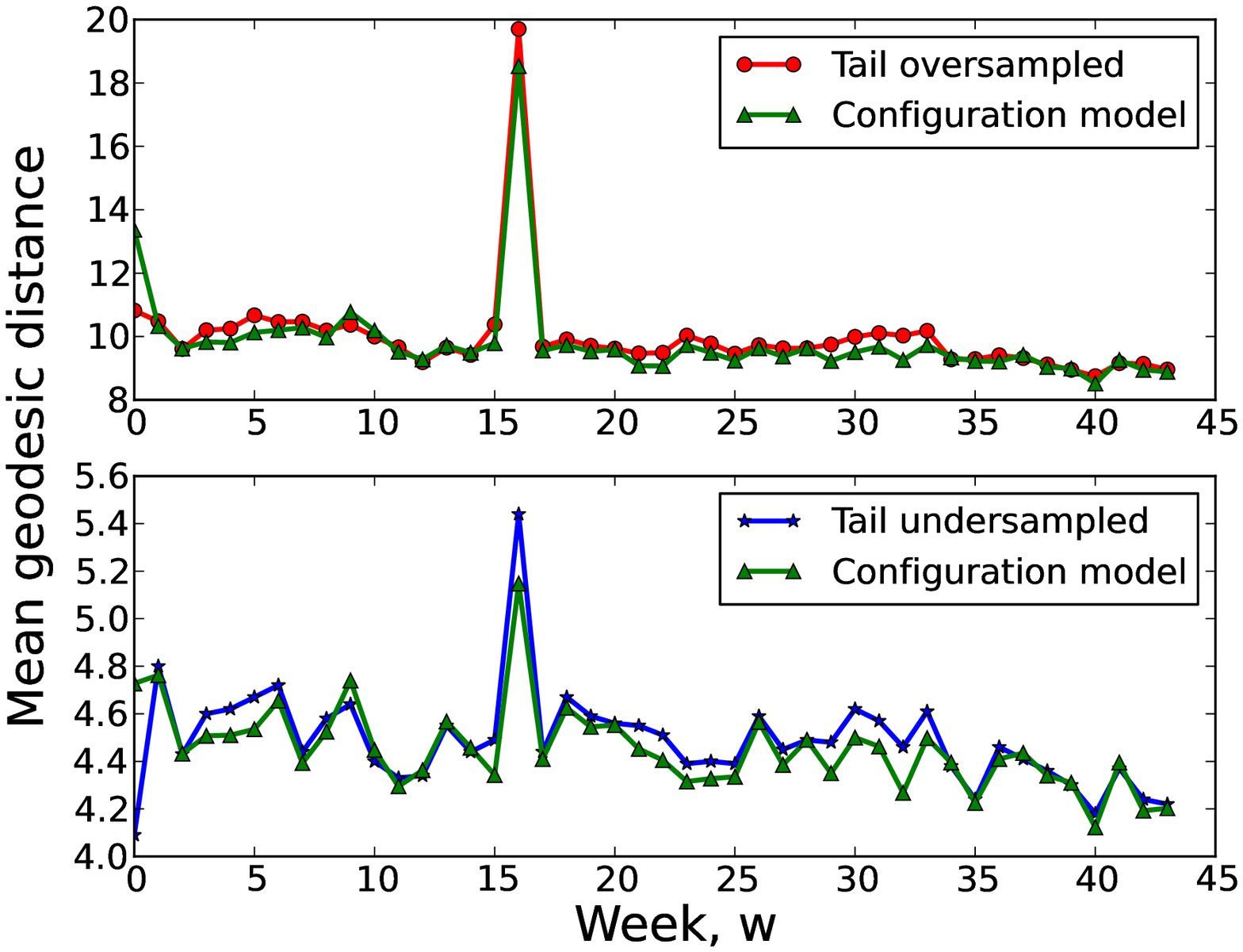} &
\includegraphics[scale=0.205]{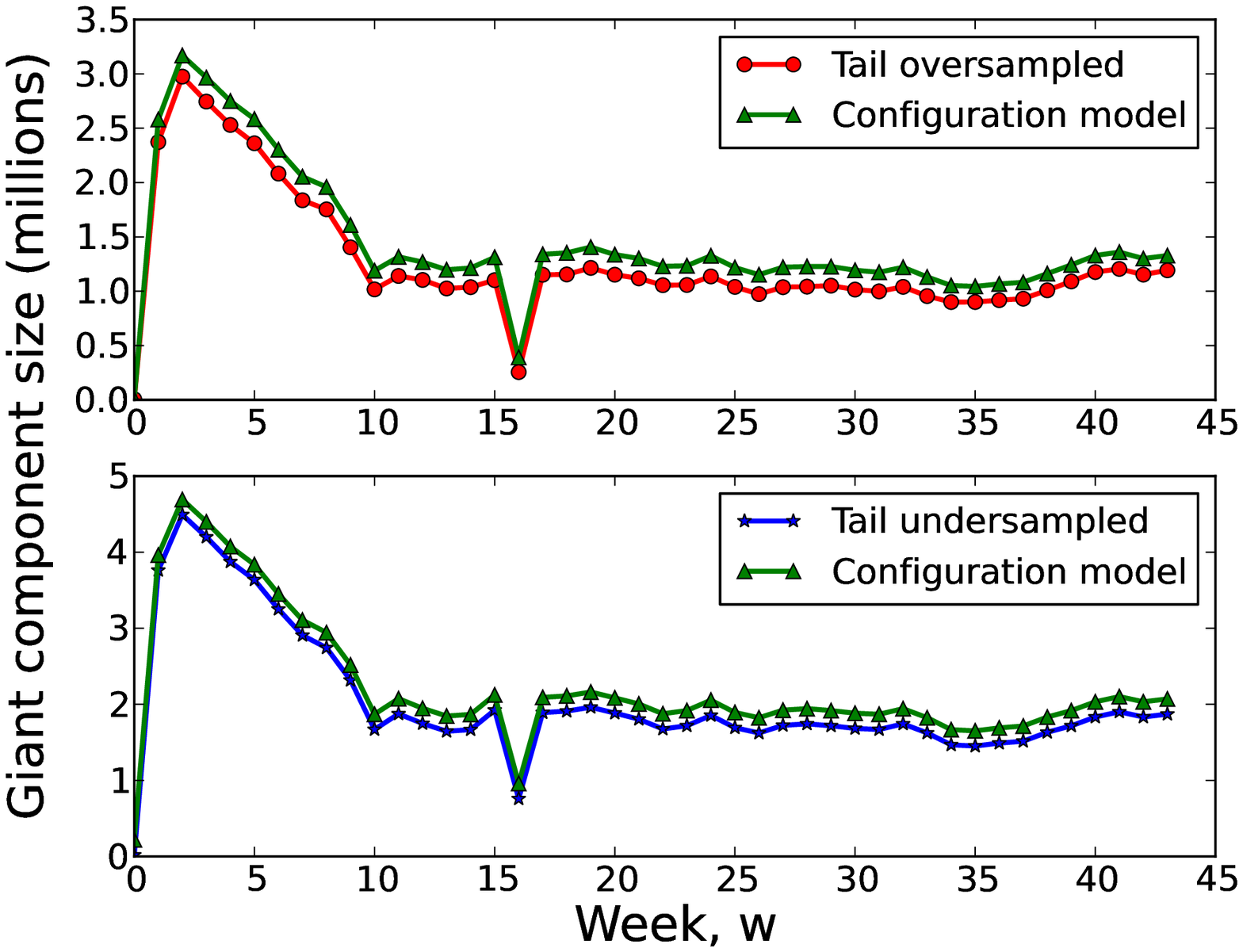} &
\includegraphics[scale=0.205]{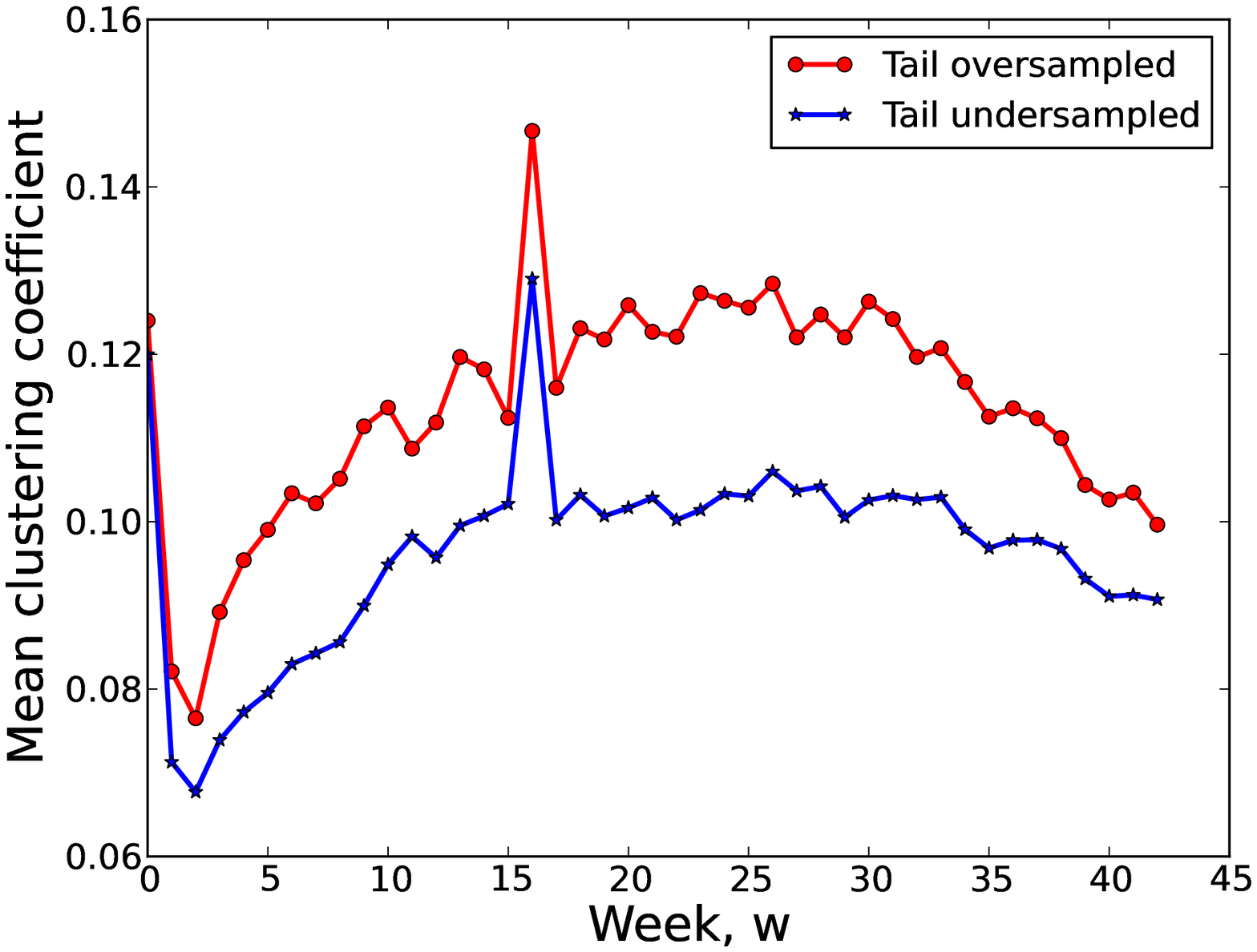} & 
\includegraphics[scale=0.205]{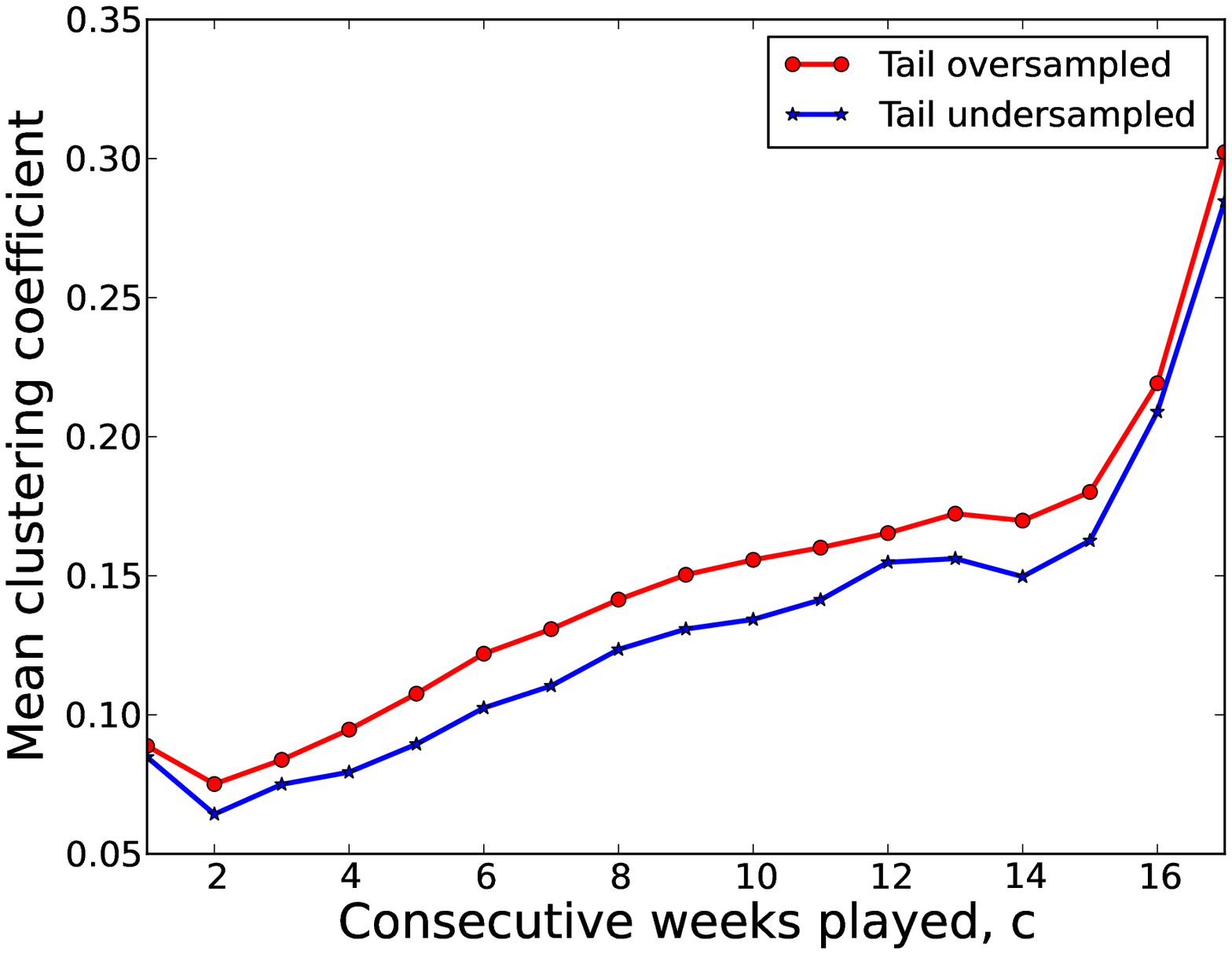}
\end{tabular}
\caption{(Left) Mean geodesic distance between 1000 randomly selected pairs of vertices in the network's giant component (Left-Center) Giant component size as a function of week, $w$. (Right-Center) Mean clustering coefficient as a function of week, $w$. (Right) Mean clustering coefficient as a function of consecutive weeks played, $c$}
\label{fig:ne:md:dist}
\end{center}
\end{figure*}

\subsection{Network non-densification}
Next, we examine how the density of the network varies with its size. In many networks, including social networks, it has been shown that a network densifies with the number of vertices~\cite{leskovec2005graphs,Kumar:2006}. That is, the number of edges in a network grows super linearly with the number of nodes. However, the friendship network studied here does not exhibit this pattern. As shown in Figure~\ref{fig:vertices:per:week}(right-center), the number of edges in the network grows linearly with the number of the vertices under both thresholds. A super linear relationship between vertices and edges would suggest that as the network grows in size, players would acquire increasing numbers of friends.

The non-densification of the network is also expressed in Figure~\ref{fig:vertices:per:week}(right), which plots the network's mean degree as a function of week, $w$. The mean degree of vertices remains relatively constant over time, with the exception of week 16, further demonstrating that the network does not densify over time. 

The linear relationship between vertices and edges in the network should come as no surprise. Establishing a link in {\em Reach} comes at a high social cost. Players must spend considerable time coordinating when and how often to play. Additionally, players must invest significant portions of time interacting with one another over the course of many games. These conditions make it difficult and time prohibitive for players to establish increasing numbers of friendships as the network grows, even though many may be available. This pattern contrasts with other online social networks, where creating an edge is as simple and cheap as accepting a friend request~\cite{Kumar:2006}.

This result indicates network densification is not a universal property of dynamic networks and that the cost of establishing links is likely an important underlying mechanism which controls the property. Thus, we should expect to find densification patterns in networks where linking is cheap and non-densification where establishing a link is expensive.

\subsection{Giant component}
Like many other social networks, the {\em Reach} friendship network is composed of many disconnected components of varying sizes (see Fig.~\ref{fig:dd:cc:cs}, right). Here, we study the dynamics and connectedness of the network's giant component, which contains between roughly 80-90\% of the network's vertices and compare it to random networks generated via the Configuration Model, using the degree sequences of each network snapshot~\cite{newman2001random}.
 
The geodesic distance between vertices is defined as the shortest path connecting them~\cite{newman2010networks}. We estimate the mean geodesic distance, also known as the average diameter, of the network at each week, $w$, by taking the mean value of the geodesic distances of 1,000 randomly selected pairs of vertices from the giant component.

As shown in Figure~\ref{fig:ne:md:dist}(left), we find the distance between vertices in the giant component to be relatively constant, with the exception of week 16, and equal to roughly 10 or 4.5 for over and under sampled tail thresholds respectively. This constant, {\em non-shrinking} diameter pattern contrasts with observations in citation~\cite{leskovec2005graphs} and other social networks~\cite{Kumar:2006}. 

The mean pair-wise distances of the friendship networks are also larger than that of the Configuration Model based random networks. This is not surprising since random graphs have no clustering. That is, in the Configuration Model networks, edges are placed between nodes at random, thereby creating a giant component with a smaller diameter, whereas in the friendship networks, edges tend to be placed within groups rather than between them, inducing a giant component with a larger diameter. Additionally, and as expected, the friendship network's clustering structure creates giant components of smaller size, when compared to the Configuration Model graphs (see Fig.~\ref{fig:ne:md:dist}, left-center). That is, the clustering structure produces more disconnected groups of vertices in the friendship network than in the Configuration Model.

\subsection{Group dynamics}
Recall, that a non-trivial fraction (between 16-20\%) of the overall population in \textit{Reach} is composed of tightly knit groups (see Fig.~\ref{fig:dd:cc:cs}, left). Here we study how clustering amongst players evolves over time.

Figure~\ref{fig:ne:md:dist}(right-center) plots the mean vertex level clustering coefficient over the course of the data's 44 week time span. During the first week, the network is clustered nearly as much as during its peak. This indicates the first week's small population of players are more social. That is, they play more in groups than the players in the immediately following weeks. After week 1, we observe a parabolic trajectory of the mean vertex level clustering coefficient. This pattern indicates that, through week 25, an increasing fraction of players tend to interact in increasingly connected groups. After this peak, the clustering within the network decreases, indicating these groups deteriorate and players choose to interact more in a pairwise fashion.

We also study how clustering correlates with play habits. For each player in the network, we calculate the largest consecutive number of weeks the player appears in the data. For each set of players playing $c$ consecutive weeks, we compute the mean clustering coefficient. 

Figure~\ref{fig:ne:md:dist}(right) plots the mean clustering coefficient as a function of consecutive weeks played and indicates a positive correlation between the number of consecutive weeks played and clustering coefficient (oversampled tail $r^2=0.85$, $p\ll0.001$, undersampled tail $r^2=0.81$, $p\ll0.001$). This pattern indicates that socially engaged players, who are members of increasingly connected groups, are retained and engaged in the system for longer and more consistent periods of time.

\section{Conclusions}
Here, we studied the evolution of a novel online social network within the popular online game \textit{Halo:\!\! Reach}, which we inferred from billions of interactions between tens of millions of individuals. We find an interesting two-phase pattern in the structural turnover of the graph, with a large ``churn'' in the beginning, as individuals try out the system briefly and then leave, followed by a more prolonged ``dynamic equilibrium'' period, characterized by a stable-sized giant component with roughly equal rates of vertices joining and leaving. Furthermore, we find that the friendship network does not densify over time and its diameter does not shrink, in contrast to other online social networks. This particular pattern is likely attributable to the high-cost of friendship links in this network, which require genuine and prolonged investment of time in order to maintain. This ``high cost'' requirement contrasts with the low- or zero-cost of maintaining links in most other online social networks. As a result, it seems likely that other online social networks with high costs for link formation and maintenance would also exhibit non-densifying patterns.

One relatively understudied aspect of online social network structure is the behavior of small groups of friends. In Halo, these groups have a functional purpose, as they constitute a coherent ``team'' of players that engage the system together. We find that the local network density (cluster coefficient) amongst individuals tends to increase over the first 25 weeks of our study period and then declines. Additionally, we find that a player's clustering coefficient is positively correlated with consecutive numbers of weeks played. This indicates players who are more socially engaged within the game are also more engaged in game play itself. 

The long term financial success of many online games, including {\em Reach}, rely not only on selling millions of copies of the game, but also on retaining players through the sale of subscription based online services. We observe that players who remain in the social network the longest are members of tighter knit groups than others. This suggests that the social aspect of {\em Reach} has a strong influence on the play patterns of its members. That is, the level of local social engagement with friends appears to correlate strongly with long-term engagement with the game itself.
One interesting question for future work is whether the overall efficacy of the game's matchmaking algorithm can be improved by explicitly accounting for the synergistic effect of playing with friends. That is, providing additional opportunities for friendship to engage socially with each other may facilitate the overall engagement of more weakly connected social groups, who may otherwise disengage the game earlier than desired.

In short, the parabolic structure of the network's clustering suggests that groups may form and disband over time. 
Studying this pattern in more detail and understanding its effects on other play patterns, such as its influence on competition outcomes and participation in different types of competitions, would shed light on how group dynamics applies to online social systems and how group skill and preferences evolve over time.

Finally, the temporal patterns observed in the Halo friendship graph demonstrate the utility of online games for studying social networks, shed new light on empirical temporal graph patterns, and clarify the claims of universality of network densification. We look forward to future studies exploring other dynamical aspects of the Halo data and the social networks embedded in other online games.


\section{Acknowledgments}
We thank Abigail Jacobs for helpful discussions, Chris Schenk for his help developing the data acquisition system and web survey, and Bungie Inc.\ for providing access to the data. This work was funded in part by the James S.\ McDonnell Foundation.

\bibliographystyle{abbrv}
\bibliography{refs}  

\balancecolumns
\end{document}